\documentclass[11pt,twoside]{article}
\usepackage{epsf}
\usepackage{asp2006}
\usepackage{graphicx}
\begin{document}

\title{Observing the Second Solar Spectrum at IRSOL}

\author{M. Bianda,$^{1,2}$ R. Ramelli,$^{1}$ and D. Gisler$^{1,2}$}   

\affil{$^1$Istituto Ricerche Solari Locarno, 6605 Locarno, Switzerland}    

\affil{$^2$Institute of Astronomy, ETH Zurich, 8093 Zurich, Switzerland}

\begin{abstract} 
The  history of the IRSOL observatory is closely related to Second Solar Spectrum observations. Already in 1963 Br\"uckner observed scattering polarization in the Ca~{\sc{i}} 4227~{\AA} line.
In 1996 the Hanle effect in the quiet chromosphere was measured for the first time in Locarno using the same spectral line.  
Since 1998 the ZIMPOL system, a polarimeter allowing unprecedented polarimetric precision, has been installed at IRSOL and been constantly upgraded to state-of-the-art technologies thanks to the close collaboration with the Institute of Astronomy in Zurich. It allows to measure the faint signatures of various scattering polarization effects. A brief historical summary of observations related to polarization at IRSOL is given.  
\end{abstract}

\section{Introduction}

In 1960 the University Observatory, USW, in G\"ottingen, Germany, opened in Locarno its station for solar observations under the earlier name {\it {Istituto per Ricerche Solari}} (Br\"uckner, Schr\"oter, \& Voigt~1967). As a consequence of the construction of the new German solar observatories on Tenerife in 1984, the stations on Capri and in Locarno were closed. A private association directed by Alessandro Rima could take over the institute in Locarno and managed to create in 1988 a foundation (FIRSOL) that governs the observatory. The name of the observatory was changed to {\it {Istituto Ricerche Solari Locarno}} (IRSOL). Since 2000 Philippe Jetzer is directing the foundation FIRSOL.
Beginning in 1988, the instrumentation has been reconstructed and upgraded with new technologies with the help of USW G\"ottingen (Wiehr \& Bianda 1994), FHS Wiesbaden  (K\"uveler, Wiehr, \& Bianda~1998, K\"uveler et al.~2003), and Institute of Astronomy at ETH Zurich. The close and fruitful collaboration with Jan  Stenflo and his Institute allowed the development of a scientific program mainly related to polarimetry within the ZIMPOL project. A historical overview of old and recent results obtained at IRSOL in this field of research is presented here, as well as some information on the ZIMPOL polarimeter itself.

\section{Second Solar Spectrum Observations in Locarno before 1984}

Under the direction of the  University Observatory in G\"ottingen, 
Germany, several projects in polarimetry were performed in Locarno. The Gregory-Coud\'e type telescope is well suited for polarization measurements. It has a low amount of instrumental polarization that remains nearly constant over the day. Thus it can be easily compensated with optical devices at the exit of the telescope or during the data analysis.

In 1963 G\"unter Br\"uckner published his work: {\it {Photoelektrische Polarisa\-tionsmessungen an Resonanzlinien im Sonnenspektrum}} (Br\"uckner 1963), reporting observations of the scattering polarization in the Ca~{\sc{i}} 4227~{\AA} line. In particular he was able to measure the  polarization in the line wings. He used a stepwise rotated half wave plate  and a calcite crystal as analyzer. The sensor was an IP21 photomultiplier.
Jan Stenflo had the opportunity to observe at Istituto per Ricerche Solari already in August 1973 and could confirm and extend the results of Br\"uckner (Stenflo 1974). Probably he did not realize at that time how familiar that place would become for him a few decades later.
Further polarimetric investigations were performed by the University Observatory in G\"ottingen.
Eberhard Wiehr (1975) measured scattering polarization in the lines Sr~{\sc{ii}}  4077~{\AA}, Ba~{\sc{i}} 4554~{\AA}, and Na~{\sc{i}} 5890~{\AA}. He also measured the center to limb variation of the continuum in several wavelength windows. His polarimeter was based on a KDP modulator (795 Hz) with which seeing effects could be eliminated.

A summary of the scientific activity performed during the first two decades when the Institute was operated by USW G\"ottingen can be found in Wiehr, Wittmann, \& W\"ohl~(1980).

\section{From two Beams Exchange Polarimeter to ZIMPOL}

The observations performed before 1984 confirmed the suitability of the Locarno telescope for polarimetric measurements. In 1995 together with Jan Stenflo and Sami Solanki it was decided to develop a two beams exchange polarimeter (Fig.~1) similar to the one used by Semel, Donati, \& Rees (1993) for stellar observations. The main goal was to perform observations in a complementary way to the first ZIMPOL version, which was not able to explore the violet-blue part of the spectrum, in contrast to the two beams exchange polarimeter. The advantages of the IRSOL two beams  exchange system are: its broad  spectral performance (limited by the calcite transparency),  its simplicity, the good polarimetric sensitivity, and its photon efficiency.
There are nevertheless limitations with this technique: two successive recordings are required to measure one Stokes parameter, and this introduces some seeing dependence in the polarization measurements. Moreover, the Stokes polarization parameters cannot be measured simultaneously, and each requires two exposures. Thus six separate exposures are needed to measure the full Stokes vector.

\begin{figure*}[!ht]
 \plotone{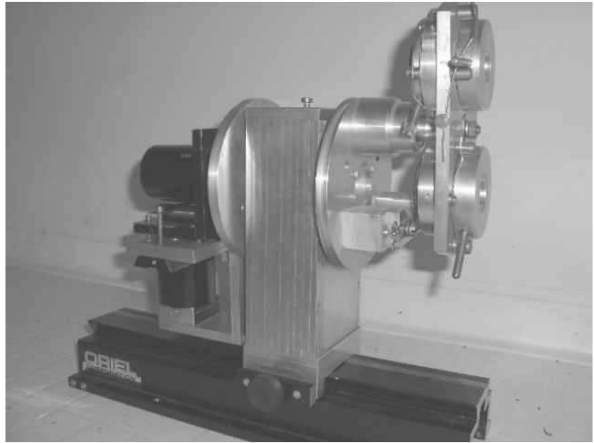}
\caption{The analyzer of the two beams exchange polarimeter, used at IRSOL during 1994-1998. Light enters from right to left. Rotating the holder on the right side, a half wave plate (top element) or a quarter wave plate (bottom element) can be inserted in the beam. Each wave plate can be rotated manually to well-defined fixed angular positions. A Savart plate inserted in the black holder, on the left side of the polarimeter, divides the incoming beam in two orthogonally polarized parallel beams.}
\label{bianda_figure1}
\end{figure*}

The main result obtained at IRSOL using the two beams exchange polarimeter was the detection of the Hanle effect in the quiet chromosphere in the lines Ca~{\sc{i}} 4227~{\AA} and Sr~{\sc{ii}} 4078~{\AA} (Bianda et al.~1998a,b). Several conclusions were presented at SPW2 in Bangalore (Bianda, Stenflo, \& Solanki~1999).
This technique is now also used at IRSOL for stellar polarimetry of bright stars (Sennhauser 2007).

In 1998 the project  to measure the {\it{Second Solar Spectrum Atlas}} with ZIMPOL~II started at IRSOL. It was the thesis work of A. Gandorfer. After a careful optimization of the instrumentation for this ambitious project, systematic observations were carried out, which resulted in the first two atlas volumes (Gandorfer 2000, 2002). Since the start of this project, a ZIMPOL system has been permanently installed at IRSOL.

\section{ZIMPOL, an Overview}

The success of IRSOL is intimately related to the Zurich IMaging POLarimeter, ZIMPOL. 
The main idea, on which this novel technology is based, was developed by Hanspeter Povel (Stenflo \& Povel~1985; Povel, Keller, \& Stenflo~1991; Povel~1995). He solved the problem how to combine polarization modulation in the kHz range (much faster than the typical seeing fluctuations) with large 2-D array detectors. 

The heart of the ZIMPOL system is the camera with a specially masked CCD sensor, in which the fast demodulation is performed. The polarization state of the incoming light is changed into temporal polarization variations by a modulator. These variations are subsequently converted into intensity fluctuations by a linear polarizer. The masked CCD has alternating open and covered pixel rows. The charges are shifted forwards and backwards between the open and covered rows in synchrony with the modulator. In this way the CCD camera can handle several simultaneous image planes corresponding to different polarization states.

The first version, ZIMPOL~I, was mainly used from 1994 to 1998 for observing campaigns at Kitt Peak, Sacramento Peak, and Tenerife. The particular CCD sensors, specially manufactured for the ZIMPOL~I cameras, have every second pixel row masked. 
The camera is synchronized with a photo-elastic modulator, PEM.
A ZIMPOL~I camera can measure Stokes $I$ and one other Stokes parameter within a single frame. To obtain the complete Stokes vector, three different cameras or sequential observations with one camera are needed. Scientific results with the system installed at Kitt Peak (Arizona) and Sunspot, Sacramento Peak (New Mexico) have been reported in a number of papers, e.g. in Stenflo \& Keller~(1997), and Stenflo et al.~(2000a,b).

A more evolved system, ZIMPOL~II, has been available since 1998.
With this system, all four Stokes parameters can be measured within one exposure. Three out of each four pixel rows of the CCD sensor are masked, allowing a single exposure to contain four images, which can be combined to reconstruct the complete polarization vector (Gandorfer \& Povel~1997). This version is efficient above 450 nm and was used at IRSOL to record the first {\it Second Solar Spectrum Atlas} (Gandorfer 2000). The ZIMPOL~II version was also used by the Institute of Astronomy in Zurich for international campaigns (Stenflo et al.~2002; Stenflo~2006).

The modulator most often used has been a PEM, with a modulation frequency of 42 kHz. To observe all four Stokes parameters with one exposure, two phase-locked PEMs would be required. So far, however, the synchronization of two PEMs within the required tolerance has not been possible. Thus PEM-modulation with ZIMPOL~II has only been possible with one PEM, which allows to measure three Stokes parameters simultaneously. To record the complete Stokes vector, two independent measurements are then needed.
The complete Stokes vector can however be recorded with one exposure with a non-resonant modulator based on two phase-locked ferro-electric liquid crystal retarders (FLC), with a typical modulation frequency of about 1 kHz. During the last couple of years ZIMPOL~II has therefore increasingly been used with a dual FLC modulation system. 

Until the year 2000 the CCD manufacturer was only able to apply a mask on frontside illuminated CCD sensors. Therefore ZIMPOL was not efficient below 450\,nm.
This wavelength limitation was overcome in 2001 by developing ZIMPOL~II-UV cameras, based on CCD sensors with an open electrode structure (Gandorfer et al.~2004). At IRSOL this camera was used to record the data needed to complete volume two of the {\it Second Solar Spectrum Atlas} (Gandorfer 2002). The third volume of this atlas (Gandorfer 2005) is based on data recorded with ZIMPOL~II-UV at McMath-Pierce Solar Telescope at Kitt Peak.

A drawback of the previous versions of ZIMPOL~II has been the low photon efficiency of the masked frontside illuminated CCD.
In 2006 a ZIMPOL~II-UV CCD sensor could be equipped with a cylindrical micro-lens array on top of the CCD surface, in order to focus all the light on the open pixel rows. The photon efficiency of the ZIMPOL~II-UV-micro-lenses camera has thereby improved by a factor of about three to six. A more detailed description of the current version of ZIMPOL~II at IRSOL can be found in Kleint, Feller, \& Bianda~(2008).

Recently a third generation of ZIMPOL has been developed (Fig.~2). It is based on state of the art electronic components, and is a very flexible system that allows instrument configurations for a wide range of polarimetric applications in astronomy. 
ZIMPOL~3 has been developed at the Institute of Astronomy in Zurich.
After the retirement of Jan Stenflo, the University of Applied Sciences of Southern Switzerland, SUPSI, in Lugano-Manno, has taken over the project. A new team is now responsible for the development and setup of ZIMPOL~3 in collaboration with IRSOL. The new system is expected to be available for science  by the end of 2008.

\begin{figure*}[!ht]
 \plotone{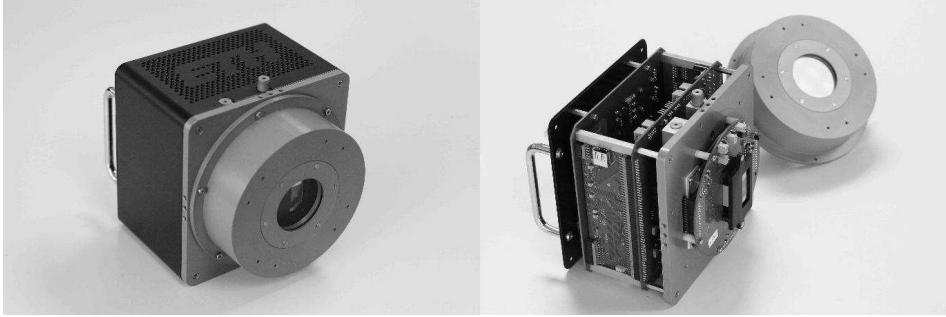}
\caption{Camera head of the ZIMPOL-3 system. The last generation of the polarimeters developed at Institute of Astronomy in Zurich is now maintained and improved at SUPSI in Lugano-Manno. The whole electronics control with the embedded control board (Linux operating system) are inside the camera box.}
\label{bianda_zimpol_img}
\end{figure*}

\section{Summary of Recent Results or Observing Programs}

Besides the already cited works performed with the two beams exchange polarimeter and the {\it{Second Solar Spectrum Atlas}}, many other observing programs have been carried out, and several of them are still ongoing. 

\subsection{Scattering Polarization Anomalies  in the Ca~{\sc{i}} 4227 \AA\ wings}

As previously mentioned, observations of the Ca~{\sc{i}} 4227~{\AA} line were already made at IRSOL in the 90s with the two beams exchange polarimeter. The availability of  ZIMPOL~II-UV allowed us to improve such observations and to discover anomalies in the linear polarization of the Ca~{\sc{i}} wings in active regions (Bianda et al.~2003a). In collaboration with K.~N. Nagendra and M. Sampoorna from IIA in Bangalore, more accurate observations were carried out, revealing that the effect is quite common and also appears in quiet regions. An example can be seen in Fig.~3, showing an observation done on 15 September 2005 at IRSOL. The usual Hanle effect signatures can be seen at line center, corresponding in $Q/I$ to local depolarization effects (darker areas at line center), in $U/I$ to Hanle rotation signatures (bright or dark areas at line center). The Hanle effect has been expected to be confined to the line core, but we can see very similar behaviour in the line wings, both in $Q/I$ and $U/I$. 
The observations and the theoretical interpretation are presented by 
Sampoorna et al.~(2009). 

\begin{figure*}[!ht]
 \plotone{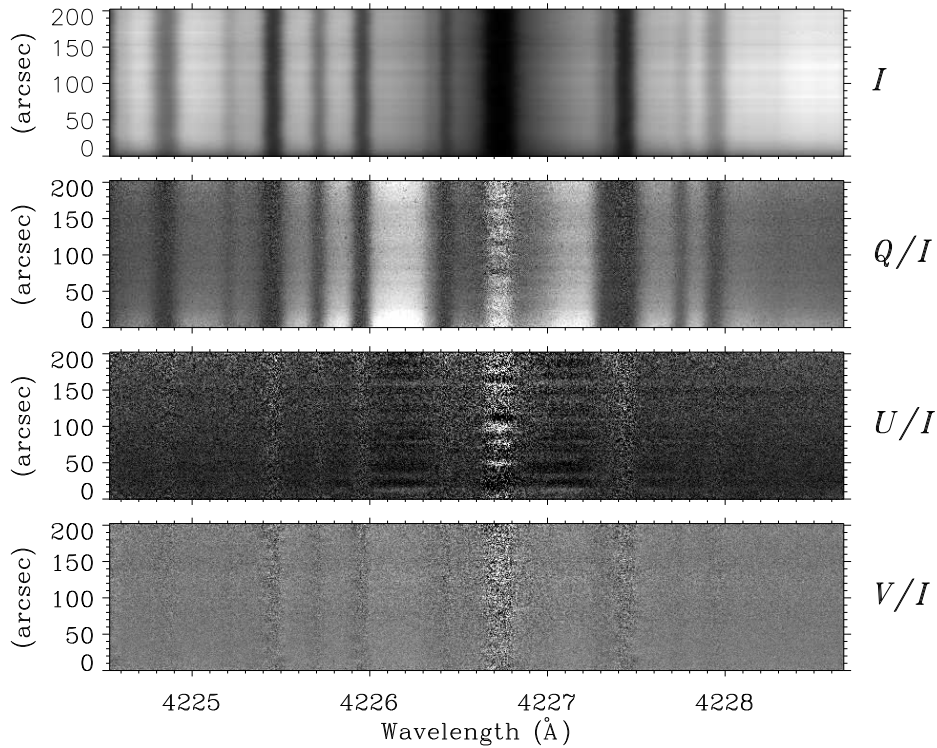}
\caption{Full Stokes vector observation of Ca~{\sc{i}} 4227 \AA\ carried out on 15 September 2005, near the W limb, at $\mu=0.1$. Note the usual Hanle signatures in $Q/I$ and $U/I$ at line center, but also note the unexpected wing signatures.}
\label{bianda_fig3}
\end{figure*}

\subsection{Absence of Impact Polarization in Solar Flares}

According to the literature (e.g. Henoux et al.~1990), most solar flares are expected to exhibit impact polarization in  H$\alpha$ in the percent regime. ZIMPOL is an ideal instrument to study this effect, because the polarization measurements are unaffected by the many instrumental effects that can produce spurious polarization signals in most other polarimeter systems. The instrumentation at IRSOL was adapted and developed to measure impact polarization with high polarimetric sensitivity and high temporal cadence (one image per second) in two-dimensional mode using an H$\alpha$ filter (FWHM 0.5 \AA). The aim was to measure impact polarization with an unprecedented polarimetric and temporal resolution. We found an upper limit to the linear polarization of about 0.1 percent for the 30 flares measured and were never able to find a significant signal, in contrast to the many claims of polarization signatures 1-2 orders of magnitude larger in the literature. To investigate the reasons for this null result we recorded tens of flares with the spectrograph, but these observations confirmed our two-dimensional null results. The results have been summarized in Bianda et al.~(2003b, 2005). Recent theoretical work supports the absence of impact polarization (e.g. {{\v S}t{\v e}p{\'a}n}, Heinzel, \& Sahal-Br{\'e}chot~2007).

\subsection{{Polarization of Prominences and Spicules}}

Spectropolarimetric observations of prominences and spicules take advantage of the reduced instrumental polarization and the low level of scattered light produced by the Gregory-Coud\'e  telescope.

First observations of prominences were done with the two beams exchange polarimeter with good quality results (Figure 4 in Wiehr \& Bianda 2003; Ramelli \& Bianda 2005). The observation method was subsequently improved, taking into account all known instrumental effects, and the measurements were extended to spicules (Ramelli et al.~2006a,b). 

The project is still ongoing, and it is planned to analyse our data using improved inversion tools (e.g. Asensio Ramos, Trujillo Bueno, \& Landi Degl'Innocenti~2008).

First observations of filaments, reported in Bianda et al.~(2006), show that it is possible to measure their faint polarization signatures.

\subsection{{Molecular Lines}}

The knowledge about molecular physics in astrophysics is increasing, and polarization of molecular lines is becoming a very important tool to study solar and stellar atmospheres, including the magnetic field. 

The Zeeman effect was studied in the G-band: observations in sunspots done at IRSOL have confirmed previous theoretical models (Asensio Ramos et al.~2004). 

CaH and TiO lines in sunspots were used to observe and study Paschen Back effects (Berdyugina et al.~2006). 

Observations with the original Semel polarimeter allowed us to study FeH lines in the IR (around 990~nm, 997~nm, and 1006~nm) in sunspots (Afram et al.~2007). 

One of the benefits of the last ZIMPOL~II version with micro-lenses is the possibility to have  good photon statistics also in the near UV. Center-to-limb variations of scattering polarization in CN molecular lines around 3870 \AA\ have been measured (Shapiro et al.~2008).

\subsection{{Second Solar Spectrum and the Solar Cycle}}

As the opportunity to observe  the Second Solar Spectrum with the required precision (RMS noise  better than $10^{-4}$) is a rather recent development, its dependence on the solar cycle is poorly known or investigated. There are already indications of variations of the scattering polarization. For example differences are seen when comparing the signatures in the Mg~{\sc{i}} triplet lines (5167~\AA, 5172~\AA, and 5184~\AA) measured by Stenflo et al.~(2000b) near the solar cycle minimum, with the Second Solar Spectrum Atlas, observed during maximum solar activity. Also other lines, like Fe~{\sc{i}} 4383.5 \AA, observed with the same instruments but during the present solar cycle minimum, are showing significant changes as compared with the atlas (Bianda, Ramelli, \& Stenflo~2007). 

\begin{figure*}[!ht]
 \plotone{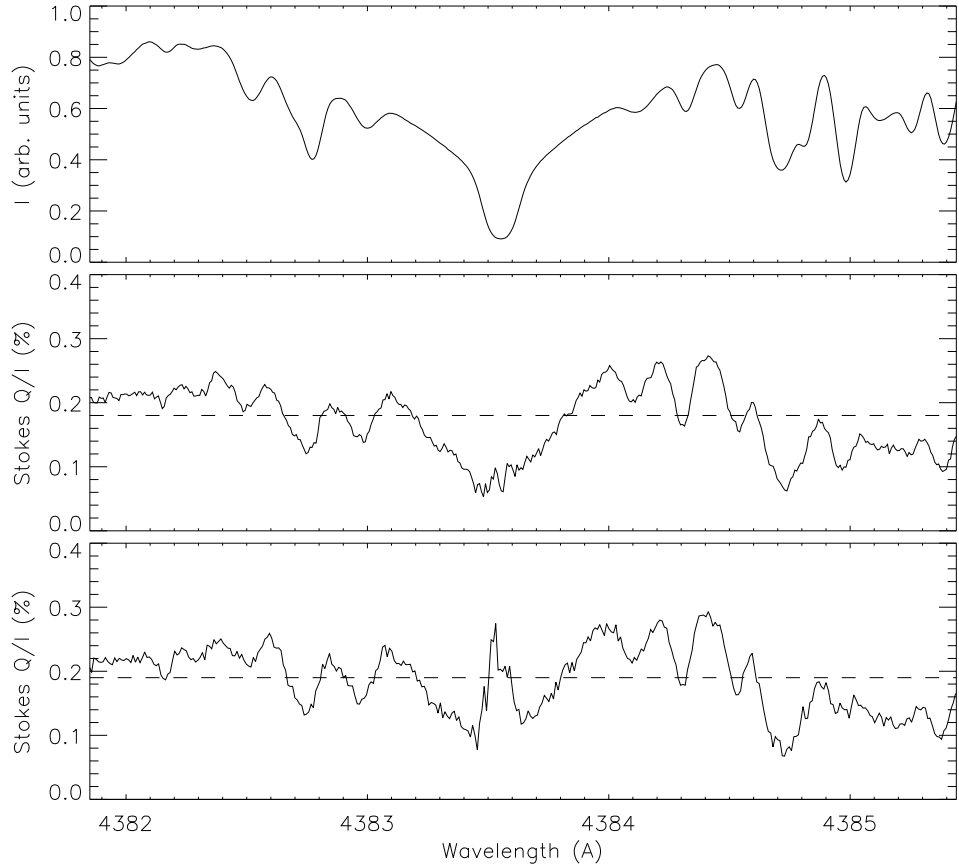}
\caption{An example of two Fe~{\sc{i}}~4383 \AA\ observations performed at $\mu=0.1$ on 10 August 2006 near the N pole (second plot) and on 23 August 2006 near the SE limb (third plot). The peak found in the {\it{Second Solar Spectrum Atlas}} (Gandorfer 2000) has an amplitude of 0.4~\%, but now smaller values are obtained. We find similar behaviour of other lines, see text.}
\label{bianda_figure4}
\end{figure*}

In Fig.~4 the Stokes $Q/I$ profiles of the Fe~{\sc{i}}~4383 \AA\ line show how the peak amplitude can change with time. Other lines are also exhibiting similar behaviour. Therefore observations are planned to better study such variations in the Second Solar Spectrum.
The two observations in Fig.~4 show that we have variations not only with 
the solar cycle, but also with heliographic latitude. Exploration of these variations gives us access to another aspect of solar magnetism that is not accessible by other means, in particular to the vast amounts of ``hidden'' magnetic fields (Trujillo Bueno, Shchukina, \& Asensio Ramos~2004; Stenflo~2004). 

Another line showing temporal variations, probably related to the solar cycle, is Na~{\sc{i}} D$_1$ 5896 \AA.  The enigmatic $Q/I$ peak reported in many publications (for example Stenflo et al.~2000a) is now found to have a smaller amplitude with variations of its shape. Further investigations are planned.

To explore the very interesting variations of the Second Solar Spectrum with latitude and time, a synoptic program has been started at IRSOL. Several lines are periodically measured to follow their evolution over a solar cycle.

\subsection{{Other Particular Observations}}

High-sensitivity spectropolarimetric observations of the D$_2$ line of 
Ba~{\sc{ii}} 4554~{\AA} (Ramelli et al.~2009) show a  good qualitative 
agreement with the theoretical model of Belluzzi, Trujillo Bueno, \& Landi Degl'Innocenti~(2007).
Spectropolarimetric measurements of this line provide a very promising tool for magnetic field diagnostics in the solar photosphere and chromosphere for a wide range of field strengths.

Observations in the Ca~{\sc{i}} 4227 \AA\ line at IRSOL, together with observations done with ZIMPOL at Kitt Peak, Arizona, have allowed a better theoretical explanation of the triplet structure of strong lines in the Second Solar Spectrum (Holzreuter, Fluri, \& Stenflo~2005). 

Sr~{\sc{i}} 4607 \AA\  center-to-limb variation data were described in \citet{bianda_stenfloetal97}, where an analytical expression was found to fit the observations. These and other ZIMPOL data were used, together with observations  by other research groups, to investigate the presence of large amounts of hidden, turbulent fields within the observed volume (Trujillo Bueno~2004; Stenflo~2004).

Taking advantage of the almost zero instrumental polarization of the IRSOL telescope at the equinox, the circular polarization of the Fe~{\sc{i}} 5247.06 \AA, 5250.22 \AA,  5250.65 \AA, and Cr~{\sc{i}} 5247.56 \AA\ lines in a sunspot observed with ZIMPOL was used to constrain the coupling constant in a theory trying to unify quantum field theory with the theory of general relativity (Solanki et al.~2004).

\section{Notes on Instrumentation}

The pointing of the telescope to a desired position is possible through the combination of different devices. The automatic guiding system (K\"uveler et al.~1998) allows us to maintain the pointing of the telescope within one arcsec. A recently developed encoder system allows precise pointing to desired coordinates within a couple of arcseconds. The image rotator coupled to the ZIMPOL modulator package allows the solar image to be rotated with respect to the spectrograph slit with an accuracy of $0.5^\circ$. Further, if observations are performed close to the solar limb (within $50\arcsec$ inside the limb), a tilt plate system developed at FHS Wiesbaden allows to compensate for image motion effects along the direction perpendicular to the limb, with an accuracy of about one arcsecond.  

An adaptive optics system, AO, similar to the infrared AO system used at Kitt Peak (Keller, Plymate, \& Ammons~2003), is developed in collaboration with SUPSI. First tests show a clear improvement of the image quality (Ramelli et al.~2006,  Figure~5). 

Besides the spectrograph observations, a two-dimensional imaging polarimetry program, based on a LiNbO$_3$ Fabry-Perot filter, has been initiated (Feller, Bianda, \& Stenflo~2007). A particular optical design that couples the filter with the spectrograph has recently been tested (Kleint et al.~2008).

IRSOL offers the opportunity to test and improve new instruments to be used in campaigns at major facilities, like at Kitt~Peak, Arizona, Sacramento~Peak, New Mexico, and Canary Islands.  A recent example is the campaign at THEMIS on Tenerife, which was organized by IRSOL. In June-July 2008 the polarimeter ZIMPOL~II with a dual FLC modulator was used by several international groups. For polarization observations at wavelengths ranging from about 450 nm to near IR, THEMIS is worldwide the best available telescope. The instrument is almost polarization free because the analyzer is inserted on the optical axis of the telescope, before the folding mirrors. The campaigns showed that the combination of ZIMPOL and THEMIS is excellent for high precision polarimetry in the visible and near IR.

\section{Concluding Remarks}

Observing with ZIMPOL at IRSOL, very high polarimetric precision can be achieved from near UV to near IR.  The instrumentation of the institute has been developed in order to optimize spectropolarimetric observations. Further performance-enhancing devices (AO, Fabry-Perot filter) are in the realization phase and will be used for scientific observations. 

The need for high quality spectropolarimetric data is increasing, also because the scientific community is now developing a number of theoretical tools to better understand Second Solar Spectrum signatures. Observations are showing new aspects that need to be explained. There is also evidence that some lines of the Second Solar Spectrum are changing with the solar cycle. There is therefore much potential for future discoveries and new physical insights about the Sun. A decade ago it was already clear that we have opened a ``new window'' in solar physics (Stenflo \& Keller 1997). 

IRSOL has had the opportunity and pleasure to be part of this fascinating development, much thanks to important collaborations like the one with Jan Stenflo and his group. Now it is time to consider new international solutions to optimize the progress of our scientific community in the further exploration of the Second Solar Spectrum. New generation telescopes (ATST, EST, balloon  and space missions) will come on line and are needed, but they lie either many years in the future, or are not always optimally designed for scattering polarization work. With already available ground-based instruments much excellent science still waits to be done. Thus there is a need to optimize the  facilities and instrumentation that already exist.

Like the  ZIMPOL campaign at THEMIS has shown, the collaborations between existing institutes can create synergies of advantage to the whole community. This is the path that we are determined to follow, with inspiration from the thought and art of Jan Stenflo. 

\acknowledgements 

Since 1988 IRSOL is managed by the foundation FIRSOL. Its president is Philippe Jetzer who succeeded  Alessandro Rima, now honorary president. The Institute is financed by Canton Ticino, ETHZ and City of Locarno together with the municipalities affiliated with CISL. Scientific projects have been supported with SNF grant 200020-117821.
Considering the historical aspect  of this article we recall the help received by USW G\"ottingen for the reconstruction of the IRSOL telescope and spectrograph, in particular by K.H. Duensing, H.H. Voigt, E. Wiehr, and A. Wittmann.
The University of Applied Sciences FHSW in Wiesbaden, Germany, collaborates since 1990 with IRSOL in the development of electronic and software devices for IRSOL under the supervision of G.~K\"uveler.
The ZIMPOL system was developed at the Institute of Astronomy in Zurich by a team composed of Frieder Abersold, Urs Egger, Stefan Hagenbuch, Peter Povel, and Peter Steiner.
The University of Applied Sciences of Southern Switzerland, SUPSI, in Lugano, collaborates with IRSOL in the development of instruments, e.g. the adaptive optics system (S. Balemi, R. Bucher), and the ZIMPOL-3 project (I. Defilippis, L. Gamma, M. Rogantini).
However, a particular acknowledgement is addressed by the personnel of IRSOL, as well as by its direction staff, to Jan Stenflo. His constructive help, his optimism, his suggestions, even in the hardest moments that IRSOL has faced, helped us to find solutions to the various problems.

\end{document}